\documentclass[twocolumn,showpacs,aps,pra]{revtex4-1}
\usepackage{ulem,bm}
\usepackage{amsmath, upgreek, amssymb, graphicx, paralist, gensymb,epstopdf}
\usepackage[colorlinks, linkcolor=blue, citecolor=blue, urlcolor=blue]{hyperref}
\usepackage{array,epsfig,amsmath,amssymb}
\usepackage{amsmath,epstopdf}
\usepackage{graphicx}
\usepackage{dsfont}
\usepackage{color}

\def\bra#1{\langle #1|}
\def\ket#1{|#1 \rangle}
\def\bracket#1#2{\langle #1|#2 \rangle}

\begin{document}

\title{
Near-deterministic quantum teleportation and resource-efficient quantum computation using linear optics and hybrid qubits}

\author{Seung-Woo Lee}
\author{Hyunseok Jeong}
\email{h.jeong37@gmail.com}

\affiliation{Center for Macroscopic Quantum Control, Department of
Physics and Astronomy, Seoul National University, Seoul, 151-742,
Korea}

\begin{abstract}
We propose a scheme to realize deterministic quantum teleportation using
linear optics and hybrid qubits.
It enables one to
efficiently perform teleportation and universal linear-optical gate operations in a simple and
near-deterministic manner using all-optical hybrid entanglement as
off-line resources.
Our analysis shows that our new approach can outperforms major
previous ones when considering both the resource requirements and
fault tolerance limits.
\end{abstract}

\maketitle

\newpage

\section{Introduction}

Quantum computers are expected to offer phenomenal increases of computational power over classical computers \cite{Nielsen2000}. There are many different approaches to implementations of quantum computers based on various physical systems while scalable quantum computation in a fault-tolerant manner is still beyond current technology. Optical models have some prominent advantages such
as relatively quick operation time compared to decoherence time \cite{PKok2007,Obrien2007,Ralph2010}. However, massive resource requirements and the gap between the fault tolerance limit and the realistic error rate should be significantly reduced \cite{Ralph2010}.

Certain properties of light can be useful to implement qubits for optical quantum information processing. Typically, photons as ``particles of light'' are considered to encode information with a well-chosen degree of freedom such as horizontal and vertical polarization states, $\ket{H}$ and $\ket{V}$. A major difficulty in this approach is to realize two-qubit gates since photons seldom interact with each other, while single-qubit operations are straightforward \cite{PKok2007,Obrien2007,Ralph2010}. In principle, scalable quantum computation can be achieved without inline nonlinear interactions \cite{Knill2001}, which is often called linear optical quantum computing (LOQC).

However, its practical implementation is difficult because LOQC gates are inherently non-deterministic \cite{Knill2001}. In the context of LOQC, quantum teleportation can be used to perform demanding two-qubit gates using the gate teleportation protocol. However, the Bell-state measurement, which forms the key element of the teleportation protocol, cannot  be performed deterministically using linear optics in this approach. Only two of the four Bell states can be identified and thus the success probability cannot exceed 50\%  \cite{Calsa2001}. It requires a very large number of
resources in order to increase the success probability of gate operations for quantum computing \cite{Knill2001} or that of the Bell-state measurement itself \cite{Grice2011}.

In general, any two distinct field states can be explored for a qubit basis \cite{Ralph2010}. Along this line, the coherent-state quantum computing (CSQC) has been developed with its own merit \cite{Jeong2001,Jeong2002,Jeong2002QIC,Ralph2003,Lund2008}. In CSQC, two coherent states, $\ket{\alpha}$ and $\ket{-\alpha}$ with amplitudes $\pm\alpha$, are used to form a qubit basis and equal superpositions of coherent states, {\it e.g.,} $\ket{\alpha}+\ket{-\alpha}$ \cite{catgen}, are required as resources \cite{Ralph2003}. Using this encoding scheme, the Bell-state measurement for coherent-state qubits ($\rm B_\alpha$) can be performed in a near-deterministic manner as $\alpha$ gets large \cite{Jeong2001}. However, a necessary single-qubit operation, {\it i.e,} $Z$-rotations, produce a cumbersome type of errors due to the non-orthogonality between $\ket{\alpha}$ and $\ket{-\alpha}$ \cite{Jeong2002,Lund2008}. This makes it still difficult to implement quantum teleporatation and gates operations in a deterministic way \cite{Lund2008}.

Towards implementations of optical quantum computation, it is important to
compare existing schemes and identify the most promising and efficient ones.
Ralph and Pryde made such a comparison \cite{Ralph2010} among major optical schemes including LOQC based on parity states (pLOQC)  \cite{Ralph2005,Hayes2010} and cluster-state approach \cite{cluster1,DEBrown2005,Dawson2006}, CSQC, and
the nonlinear Zeno protocol \cite{zeno1,zeno2}.
They identified pLOQC and CSQC as the best ones
when considering both the loss threshold for fault-tolerant quantum computing and the resource requirements \cite{Ralph2010}.

In this paper, we devise an approach based on
all-optical hybrid qubits devised to combine advantages of
LOQC and CSQC.
In particular, we show that near-deterministic quantum teleportation
can be performed using linear optics and hybrid qubits.
Our approach enables one to perform near-deterministic universal gate operations
for efficient scalable quantum computation.
Remarkably, it outperforms LOQC and CSQC when resource
requirements  and error thresholds are considered together.
Our work thus paves an efficient new way for the optical realization of practical quantum computation.

\section{Deterministic quantum teleportation and universal gate operations using hybrid qubits}

\subsection{Hybrid optical qubits and single qubit operations}

In our approach, the orthonormal basis to define optical hybrid qubits is
\begin{eqnarray}
\nonumber \big\{\ket{0_L}=\ket{+}\ket{\alpha},~~
\ket{1_L}=\ket{-}\ket{-\alpha}\big\},
\end{eqnarray}
where $\ket{\pm}=(\ket{H}\pm\ket{V})/\sqrt{2}$
and $\alpha$ is assumed to be real without loosing generality.
As we shall see, this approach enables us to overcome particular weak points of both LOQC and CSQC at the same time. The $Z$-basis measurement can be performed by a single measurement on either of the two physical modes. It can be done on the single-photon mode by a polarization measurement on the basis $\ket{+}$ and $\ket{-}$, or on the
coherent-state mode using an ancillary coherent state \cite{Jeong2002}.

In our scheme, the Pauli $X$ operation, $\hat{X}$, can be performed by applying a bit flip operation on each of the two modes. The bit flip operation on the single-photon mode, $\ket{+} \leftrightarrow \ket{-}$, is implemented by a polarization rotator, and the operation on the coherent-state mode, $\ket{\alpha} \leftrightarrow \ket{-\alpha}$, by a $\pi$ phase shifter. An arbitrary $Z$ rotation ($\hat{Z}_{\theta}$) is performed by applying the phase shift operation only on the single-photon mode: $\{\ket{+},~\ket{-}\} \rightarrow \{\ket{+},~e^{i\theta}\ket{-}\}$, and no operation is required on the coherent-state mode. This is a significant advantage over CSQC in which $Z$ rotations are highly nontrivial and cause a heavy increase of the circuit complexity \cite{Lund2008}.

\subsection{Resource states for universal gate operations}

In order to construct a universal set of gate operations, Pauli $X$, arbitrary $Z$ (phase) rotation, Hadamard, and controlled-$Z$ (CZ) gates suffice \cite{Nielsen2000}.
In our scheme, the necessary resource states for universal gate operations are
the {\em hybrid pairs}, $\ket{H}\ket{\alpha}+\ket{V}\ket{-\alpha}$.
We shall also present an alternative method using both the
{\em two-photon pairs}, $\ket{H}\ket{H}+\ket{V}\ket{V}$, and the hybrid pairs.
As we shall discuss, each of the two methods has its own merit, but overall the method using only hybrid pairs
shows the better performance.
A hybrid pair can be generated in principle by performing a weak cross-Kerr nonlinear interaction between a single photon and a strong coherent state together with a displacement operation \cite{Nemoto2004,Jeong2005,Munro2005}. It has been shown that a high-fidelity cross-Kerr nonlinearity can be obtained \cite{BHe2011,Hosseini2011,Chudzick2012} despite a limitation in optical fibers \cite{Jeff2006,Jeff2007}. As we shall see, we only need small-scale hybrid pairs ({\it e.g.} $\alpha\approx 1$) of which the demonstration using gradient echo memory \cite{Hosseini2011} would be experimentally feasible in the foreseeable future.

\subsection{Near-deterministic quantum teleportation using linear optics}

\begin{figure}
\includegraphics[width=0.9\linewidth]{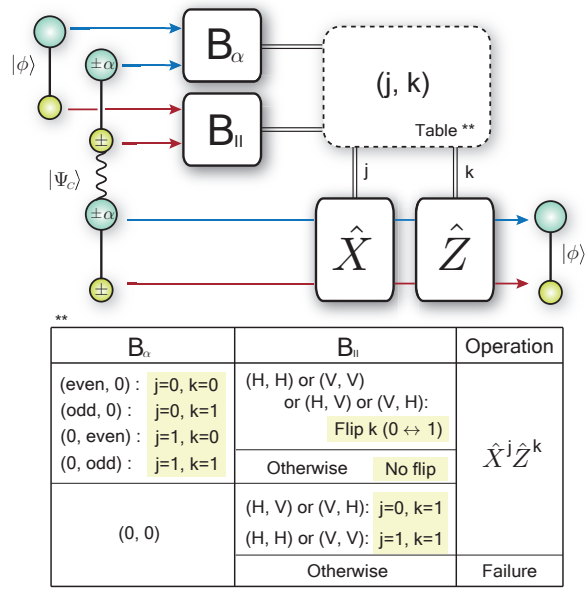}
\caption{(Color online) Scheme for near-deterministic quantum
teleportation for a hybrid qubit using linear optics and photon detection. An unknown hybrid qubit,
$\ket{\phi}=a\ket{0_L}+b\ket{1_L}$, is teleported through channel
$\ket{\Psi_C}\propto \ket{0_L}\ket{0_L}+\ket{1_L}\ket{1_L}$. $\rm B_{\alpha}$ and $\rm B_{II}$ are performed on
coherent-state modes and photon modes, respectively, between the
qubit and one party of the channel state. All possible outcomes and
corresponding feed-forward operations are presented in the table. A
failure occurs when both $\rm B_{\alpha}$ and $\rm B_{II}$ fail. The
failure probability is found to be $P_{f}=\exp(-2\alpha^2)/2$. In
order to perform Hadamard and CZ gates, entangled states $\ket{Z}$
and $\ket{Z^\prime}$ should be used, respectively, instead of
$|\Psi_C\rangle$. }\label{fig1}
\end{figure}

A teleportation protocol is required to perform Hadamard and CZ
operations \cite{Gottesman1999}. Using our approach, teleportation can be performed in a simple and near-deterministic manner. We emphasize that this is an extremely difficult task in the framework of LOQC because of the limited
success probability of the Bell state measurement using linear optics up to 50\% \cite{Calsa2001} or the required large number of modes prepared in single photon states for high-success teleporter \cite{Knill2001}. It is also difficult in CSQC due to the difficulty in performing deterministic $Z$ rotations, which is the cost of using a non-orthogonal qubit basis \cite{Lund2008}.

In our teleportation scheme, the Bell measurement for an optical hybrid qubit
can be performed using two smaller Bell measurement units  as shown in Fig.~\ref{fig1}.
A coherent-state Bell measurement, $\rm B_{\alpha}$, is implemented
by a 50:50 beam splitter and two photon number parity detectors
(PNPDs) \cite{Jeong2001}.
It unambiguously discriminates between all four coherent-Bell states and the success
probability is $1-\exp(-2\alpha^2)$ \cite{Jeong2001}.
A non-deterministic Bell
measurement or type-II fusion operation \cite{Calsa2001}, $\rm
B_{II}$, identifies only two of the Bell states, {\it e.g.},
$(\ket{H}\ket{V}\pm\ket{V}\ket{H})/\sqrt{2}$,
using four on/off photodetectors with success probability $1/2$
(Details of $\rm B_{\alpha}$ and $\rm
B_{II}$ are reviewed in Appendix A).

Suppose that an unknown hybrid qubit, $\ket{\phi}=a\ket{0_L}+b\ket{1_L}$, is to be teleported using
entangled channel $\ket{\Psi_C}\propto \ket{0_L}\ket{0_L}+\ket{1_L}\ket{1_L}$.
The smaller Bell measurements, $\rm B_\alpha$ and $\rm B_{II}$, are performed in each mode together with one part of the channel $\ket{\Psi_C}$ as depicted in Fig.~\ref{fig1}. According to the measurement results, appropriate Pauli operations are determined as shown in the table of Fig.~\ref{fig1} which completes the teleportation process.
For example, if
 the ``upper''  detector of $\rm B_\alpha$ (that employs two PNPDs) detects an odd number of photons
 while the ``lower'' one  does not click,
the outcome of $\rm B_\alpha$ is (odd, 0) and
we assign $j=0$ and $k=1$ as shown in the table of  of Fig.~\ref{fig1}.
At the same time, in the $\rm B_{II}$ measurement that uses four on/off detectors, if one detector among the upper two {\it and} another from the lower two click, this means that the outcome
 is $(H, H)$ or $(H, V)$ or $(V, H)$ or $(V, V)$ in
Fig.~4 of Appendix A \cite{Calsa2001}. In this case, we flip the assigned values
as described in the table  so that $j=1$ and $k=0$
are obtained. (Otherwise, $j$ and $k$ remain unchanged.)
 Finally, the feed-forward operation $\hat{X}^{j}\hat{Z}^{k}$ on the output hybrid qubit in the channel completes the teleportation.

  The process will be successful unless both $\rm B_\alpha$ and $\rm B_{II}$ fail;
 even though the $\rm B_\alpha$ fails, the input state can be fully teleported if $\rm B_{II}$ is successful
as shown in the table of Fig.~\ref{fig1}.
    This leads to the failure probability of
\begin{equation}
P_f=\frac{1}{2}e^{-2\alpha^2},
\label{eq:fail}
\end{equation}
 which outperforms the previous schemes that require massive overheads with repetitive applications of teleporters \cite{Knill2001,Lund2008}. For example, $99\%$ success rate of teleportation is achieved by encoding with $\alpha=1.4$.

Of course, a maximally entangled state, $\ket{\Psi_C}$, in the hybrid basis is required a quantum channel for teleportation. It can be generated for example by combining a hybrid pair of amplitude $\sqrt{2}\alpha$ and a two-photon pair by $\rm B_I$ as shown in Fig.~\ref{fig2}(a).
A state in the form of
$|H\rangle|\alpha\rangle|\alpha\rangle+|V\rangle|-\alpha\rangle|-\alpha\rangle$
is obtained using a 50:50 beam splitter from the hybrid pair of amplitude $\sqrt{2}\alpha$. The $\rm B_I$ operation
then combines it with the two-photon pair so as to generate $|\Psi_C\rangle$.
Detailed analysis and an alternative generation scheme are introduced in Appendix B.

\subsection{Hadamard and CZ operations}

\begin{figure}
\begin{center}
\includegraphics[width=1\linewidth]{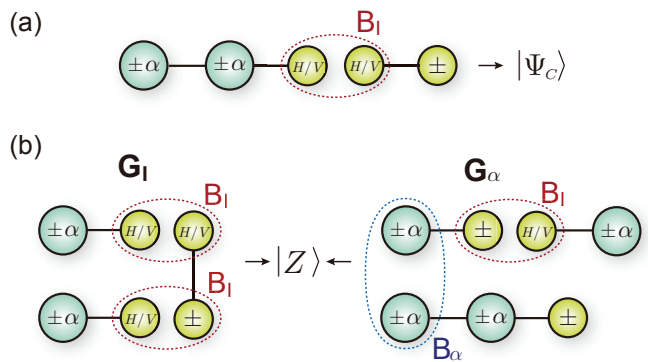}
\caption{(Color online) Schemes to prepare entangled channels. (a) A
maximally entangled state, $|\Psi_C\rangle$, is generated using a
$\rm B_I$ operation out of a two-photon pair and a hybrid pair of
$\sqrt{2}\alpha$. In a single-photon mode, $\pm$ and $H/V$ denote
bases $\{\ket{+},\ket{-}\}$ and $\{\ket{H}$,$\ket{V}\}$,
respectively, which can be modified by a polarization rotation
before performing $\rm B_I$ operation. A 50:50 beam splitter, used
to split the coherent-state part with amplitude $\sqrt{2}\alpha$
into two modes, is omitted in the figure. (b) Two schemes, $\rm G_I$
and $\rm G_\alpha$, to generate $|Z\rangle$. In $\rm G_I$, two $\rm
B_I$ operations are performed on one two-photon pair and two hybrid
pairs so as to link them. The other method, $\rm G_\alpha$, requires
three hybrid pairs with $\rm B_I$ and $\rm B_\alpha$ operations. One
of those hybrid pairs has amplitude $\sqrt{2}\alpha$ to obtain a
three-mode state
$|+\rangle|\alpha\rangle|\alpha\rangle+|-\rangle|-\alpha\rangle|-\alpha\rangle$
by a 50:50 beam splitter (omitted in the figure). Appropriate
feed-forwards with Pauli-operations are necessary for all $\rm B_I$
and $\rm B_\alpha$ operations dependent on the measurement outcome.}\label{fig2}
\end{center}
\end{figure}

In order to perform the Hadamard and CZ gates, entangled states
$|Z\rangle\propto\ket{0_L}\ket{0_L}+\ket{0_L}\ket{1_L}+\ket{1_L}\ket{0_L}-\ket{1_L}\ket{1_L}$
and $|Z^\prime\rangle\propto\ket{0000}+\ket{0011}+
\ket{1100}-\ket{1111}$ where $\ket{0000}=\ket{0_L}\ket{0_L}\ket{0_L}\ket{0_L}$ and so on, should be used as the teleportation channel, respectively.
We present  two different schemes, $\rm G_I$
and $\rm G_\alpha$, to generate $|Z\rangle$
as shown in Fig.~\ref{fig2}(b).
Using $\rm G_I$, two $\rm
B_I$ operations are performed on one two-photon pair and two hybrid
pairs so as to link them. The other method, $\rm G_\alpha$, requires
three hybrid pairs with $\rm B_I$ and $\rm B_\alpha$ operations. One
of those hybrid pairs has amplitude $\sqrt{2}\alpha$ to obtain a
three-mode state
$|+\rangle|\alpha\rangle|\alpha\rangle+|-\rangle|-\alpha\rangle|-\alpha\rangle$
by a 50:50 beam splitter. Appropriate
feed-forwards with Pauli-operations are necessary for all $\rm B_I$
and $\rm B_\alpha$ operations dependent on the measurement outcome.
 The four-qubit entangled state, $|Z^\prime\rangle$, can also be generated in a similar manner with about twice of the resources using either $\rm G_I$ or $\rm
G_\alpha$.
It should be noted that only hybrid pairs are required when $\rm G_\alpha$ is chosen as the generation strategy,
while  both two photon pairs and hybrid pairs are required when using $\rm G_I$.
Details of all generation schemes of entangled states are presented in Appendix B. We emphasize that these states are prepared as off-line resources while linear optical elements with photon detections are sufficient for inline operations.

\section{Performance analysis for fault-tolerant and scalable quantum computation}

\subsection{Errors analysis}

 Errors due to photon losses are considered a major detrimental
factor in optical quantum computing \cite{Ralph2010}. Some errors
are immediately noticed during gate operations, which are called
locatable errors \cite{Ralph2010}.
 Unlocatable errors are detectable
only with an error correcting code. Losses at single-photon modes
are locatable by $\rm B_{II}$ whenever performing teleportation for
Hadamard or CZ gates. Furthermore, a missing photon at a single-photon mode
is immediately compensated in the output qubit $|\phi\rangle$ as far
as $\rm B_{\alpha}$ succeeds as clearly seen in Fig.~\ref{fig1}.
However, losses at coherent-state modes may cause unlocatable errors
besides locatable ones. This is due to the fact that a coherent state does not contain
a definite number of photons so that is it cannot be noticed when a photon is lost.

We analyze locatable and unlocatable errors with
loss rate $\eta$.
The analytical solution of the hybrid qubit and error rates under loss effects
can be obtained using  the master equation, and the full results are presented  in Appendix C.
Under the loss effects,
the failure probability $P_{f}$ for teleportation in
Eq.~(\ref{eq:fail})  is modified to
\begin{equation}
P'_{f} =
(1-\eta)\frac{e^{-2\alpha'^2}}{2}+\eta\frac{2}{1+e^{2\alpha'^2}}
\end{equation}
where $\alpha'=\sqrt{1-\eta}\alpha$. If
a gate operation fails, the teleported qubit is assumed to
experience depolarization and become fully mixed. This is equivalent
to applying a random Pauli operation to the qubit, {\it i.e.} $Z$
and $X$ Pauli errors occurs independently with the equal
probabilities. One can also assume that if a loss occurs in either
photon or coherent-state modes, the hybrid qubit experiences a Pauli
$Z$ error with probability $1/2$. We also model errors due to losses
in the generation processes of $|Z\rangle$ and $|Z'\rangle$ as Pauli
$X$ and $Z$ errors. We assume that the decrease in amplitude
$\alpha$ by loss can be compensated whenever using the teleportation
scheme by changing the amplitude of output state of the channel
\cite{Lund2008}. Based on these models and methods, we have
analytically obtained probabilities of aforementioned errors in
terms of $\eta$ (Appendix C).

In order to realize scalable quantum computation, it should be justified that arbitrarily large computation can be implemented with small errors, which called to be fault tolerance \cite{Shor1996}. In this sense, a fault tolerant noise threshold can be obtained such that if the amount of noise per operation is below this threshold, it is possible to realize an arbitrary large scale quantum computers with appropriate error corrections
\cite{Nielsen2000,Knill2005,Steane2003}.

We employ an error correction protocol with several levels of concatenation based on the circuit-based telecorrection \cite{Dawson2006}. Using the telecorrection protocol \cite{Dawson2006}, noise thresholds and resource requirements in cluster-state LOQC \cite{Dawson2006}, pLOQC \cite{Hayes2010}, and CSQC \cite{Lund2008} were previously investigated. In order to compare our approach with the previous ones, we follow the same analysis using the 7-qubit STEANE code \cite{Steane1996} based on the telecorrection protocol. We assume our error model for the lowest level of concatenation. For higher levels, the noise model and error-correction protocol are identical to those of Ref.~\cite{Dawson2006}. We perform a numerical simulation (Monte Carlo method using C++) for one round of the error correction for the first level concatenation. The modified telecorrector circuit is composed of CZ, Hadamard, $\ket{+}$ creation, and $X$-basis measurement \cite{Lund2008}.

We carried out a series of simulations for a range of loss rate
$\eta$ and amplitude $\alpha$. The resulting rates of unlocatable
and locatable errors are used for the next level of concatenation
for the error correction. If the error rates tend to zero with
certain values of $\eta$ and $\alpha$ in the limit of many levels of
concatenation, fault-tolerant computing is possible with those
values. In this way, the noise threshold curves are obtained.

\subsection{Resource requirements}

Once a fault tolerant model is determined, the number of resources, required for one round of error correcting may be considered as another crucial factor for scalability. We consider two-photon pairs and hybrid pairs to be resources. We estimate the number of resources required for one round of error correction in the lowest level of concatenation. It is assumed, following the estimation in Refs.~\cite{Lund2008,Ralph2010}, that the total number of operations in one round of the telecorrection scheme is about 1,000 \cite{Ralph2010} and resources are used in each operation as the following fractions \cite{Lund2008}: Memory 0.284, Hadamard 0.098, CZ 0.343, Diagonal state (hybrid-pair) 0.164, and X-basis measurement 0.111. We also assume parallel productions of resource states and no reuse of resources to avoid complicated techniques used for saving resources.

We have two types of generation schemes, $\rm G_I$ and $\rm G_{\alpha}$, which consume different numbers of resources to produce $\ket{Z}$ and $\ket{Z'}$. When generating $\ket{Z}$ by $\rm G_I$, two $\rm B_I$ operations (the success probability of each operation is 1/2) are used to merge one two-photon pair
and two hybrid pairs ({\it i.e.,} 3 resources). Since the success probability of the two $\rm B_I$ operations is $1/4$, the average number of required resources is $3\times 4=12$ ({\em i.e.}, 4 two-photon and 8 hybrid pairs on average are required). In $\rm G_{\alpha}$, on the other hand, $\rm B_I$ and $\rm B_{\alpha}$ are performed once each to merge three hybrid pairs. The success probability is then $(1-e^{-2\alpha^2})/2$ because the failure probabilities of $\rm B_\alpha$ and $\rm B_{II}$ is $e^{-2\alpha^2}$ and $1/2$, respectively, as discussed in Sec.~II. Thus, $6/(1-e^{-2\alpha^2})$ number of hybrid pairs are required in average. Likewise, when generating $\ket{Z'}$, 48 two-photon pair and 32 hybrid pairs (80 resource states) are used in $\rm G_I$,
while $10/(1-e^{-2\alpha^2})^{3}$ number of hybrid pairs are required in $\rm G_{\alpha}$.

Therefore, the total number of resources required for one round of error correction is obtained as $98\times12+343\times80+164=28780$ for $\rm G_I$, while it is $98\lceil 6/(1-e^{-2\alpha^2})\rceil+343\lceil 10/(1-e^{-2\alpha^2})^{3}\rceil+164$ for $\rm G_{\alpha}$ which becomes down to $4182$ as increasing $\alpha$.

\begin{figure*}
\includegraphics[width=1\linewidth]{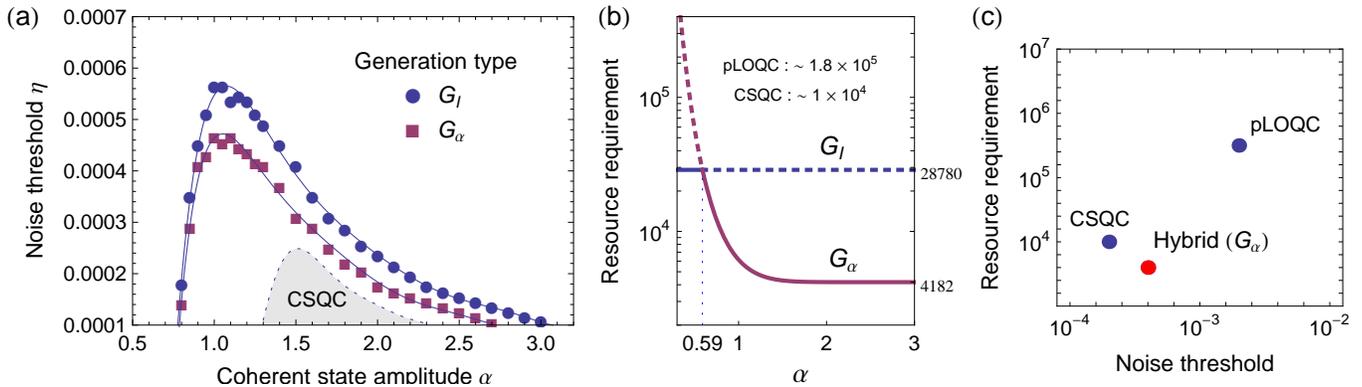}
\caption{(Color online) Noise thresholds and resource requirements.
(a) Noise thresholds based on two generation schemes, $\rm G_I$ and
$\rm G_\alpha$, obtained using the 7-qubit STEANE code
\cite{Steane1996}. (b) Resource requirements estimated for one round
of error correction based on the telecorrection protocol. For $\rm
G_I$, a constant number (28780) of resources (two-photon and hybrid
pairs) are required irrespective of amplitude $\alpha$, while for
$\rm G_\alpha$ the required number of resource states tend to
decrease rapidly down to 4182 as increasing $\alpha$. Two curves
intersect at $\alpha \approx 0.59$. We can take the lower curve
between them by choosing  $\rm G_\alpha$ for $\alpha > 0.59$ and
otherwise  $\rm G_I$. It shows a remarkable improvement compared to
pLOQC (about $1.8\times10^5$ \cite{Hayes2010}) and CSQC (about
$10^4$ \cite{Lund2008}).
(c) Ralph-Pryde diagram \cite{Ralph2010} for the comparison with pLOQC and CSQC.
The hybrid approach using $\rm G_\alpha$ presented in this paper is shown to
outperform pLOQC and CSQC.
   }\label{fig3}
\end{figure*}

\subsection{Comparison with LOQC and CSQC}

Ralph and Pryde  suggest that LOQC and CSQC are the best schemes
for medium scale quantum computing
considering both the fault-tolerant thresholds and the resource costs
\cite{Ralph2010}.
We here present the results of our numerical analysis in order to compare the performance of our scheme, when it is applied to scalable quantum computing, with  LOQC and CSQC.
As shown in Fig.~\ref{fig3}(a) for $\rm G_I$ and $\rm G_\alpha$, the noise threshold level is obviously higher than CSQC for all region of $\alpha$ while it is still lower than that of pLOQC (about $2\times10^{-3}$ \cite{Hayes2010}).
The threshold peak for each generation scheme appears around
$\alpha \approx 1.08$ (Fig.~\ref{fig3}(a)). However, further increase of $\alpha$ lowers the threshold level due to rapid increase of unlocatable errors which are more difficult to correct than locatable ones using the telecorrection protocol \cite{Dawson2006}. The noise thresholds of $\rm G_I$ appear to be slightly larger than those of $\rm G_\alpha$ because $\rm G_\alpha$ requires preparation of hybrid qubits with amplitude $\sqrt{2}\alpha$ from the beginning as seen in Fig.~\ref{fig2}.

Remarkably, our scheme provides a greatly reduced resource cost compared to both CSQC and pLOQC.
This is partly due to the near-deterministic nature of our teleportation protocol.
As presented in Fig.~\ref{fig3}(b), resource requirements are phenomenally reduced by $\rm G_\alpha$ since the success rate of $\rm B_{\alpha}$ grows rapidly as $\alpha$ increases.
However, the success rate of $\rm B_I$ is constant (1/2) and so are the resource requirements with $\rm G_I$.
The diagram in Figure~\ref{fig3}(c) clearly shows that our scheme (with $\rm G_\alpha$) shows better performance
over pLOQC and LOQC when both the resource cost and the noise threshold are considered.


\section{Remarks}

In this paper, we have developed an all-optical hybrid scheme of quantum computation. We have shown that near-deterministic quantum teleportation can be performed using linear optics and hybrid qubits. Our approach enables one to perform near-deterministic universal gate operations for efficient scalable quantum computation. It was shown to outperform major previous ones when resource requirements and error thresholds are considered together. The required {\it offline} resources states  are
hybrid pairs in the form of $|H\rangle|\alpha\rangle+|V\rangle|-\alpha\rangle$
and only a small value of the amplitude as $\alpha\approx1$ are required to demonstrate the maximum performance.

Towards fault-tolerant and scalable quantum computation, a crucial experimental challenge would be to enhance the efficiencies of the photon detectors. In fact, efficiencies of currently available detectors \cite{Eisaman2011} are far from the required levels to overcome the fault tolerant limits. This is a critical problem in any type of optical approaches to quantum computing (including LOQC and CSQC). Our scheme requires, as CSQC does, photon number resolving detectors for parity measurements and this is one reason that its fault-tolerant limit is still lower than pLOQC
(but higher than CSQC).

Efficient preparation of the resource hybrid pairs with high fidelities along with the current progress of optical controls \cite{Ralph2010} would be the next challenge in the development of our scheme. It is known that the fidelities of hybrid pairs generated using weak nonlinearities in optical fibers are limited \cite{Jeff2006,Jeff2007}.
Since we only need small-scale hybrid pairs ({\it e.g.} $\alpha\approx 1$),
their high-fidelity generation may be possible using gradient echo memory \cite{Hosseini2011}.

One may also consider possibility that there exist more than one photon
in the single-photon part of a hybrid pair (or one part of a two-photon pair)
due to experimental imperfections  \cite{Kwiat1995}.
There effects may be significantly suppressed by the $\rm B_I$  and $\rm B_{II}$ operations
during the generation and inline processes.
(see Appendix D for a detailed discussion).

Given the deterministic nature of our scheme and its performance over the previous ones,
we expect that our work will pave an efficient way for the optical realization of scalable quantum computation.
There exist experimental obstacles such as highly efficient detectors and high-fidelity resource states towards realizations of scalable quantum computation.
In fact, the gaps between fault-tolerance limits and efficiencies of currently available on/off
detectors and photon number resolving detectors are still formidable \cite{Eisaman2011}.
On the other hand, demonstration of the teleportation
scheme for a hybrid qubit would be experimentally feasible in foreseeable future.

\acknowledgments

This work was supported by the National Research Foundation of Korea (NRF)
grant funded by the Korean Government
(No. 3348-20100018)
and the World Class University program.

\renewcommand{\theequation}{A-\arabic{equation}}
\setcounter{equation}{0}
\section*{Appendix A: Review of Bell-type measurements}
\label{Appendix1}

\begin{figure}
\begin{center}
\includegraphics[width=0.9\linewidth]{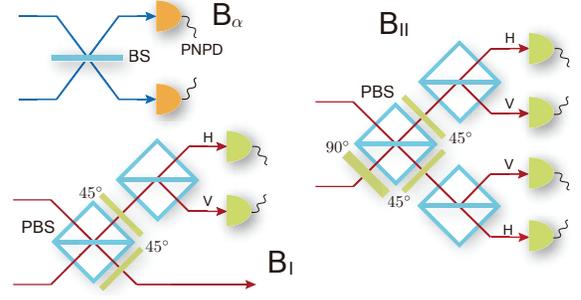}
\caption{{\bf Three Bell-type measurement elements used for our
scheme.} A coherent-state Bell measurement, $\rm B_{\alpha}$, is
implemented by a 50:50 BS and two photon number parity detectors
(PNPD) \cite{Jeong2001}. $\rm B_{\alpha}$ unambiguously
discriminates between all four Bell states and the success
probability is $1-\exp(-2|\alpha|^2)$. It fails only when no photon
is detected at both the detectors. A type-I fusion operation
\cite{DEBrown2005}, $\rm B_{I}$, is implemented by polarizing beam
splitters (PBS), wave plates, and photon detectors. It effectively
performs $\ket{+}\bra{H}\bra{H}\pm\ket{-}\bra{V}\bra{V}$ with a
success probability $1/2$ when only one photon is detected at either
detectors \cite{DEBrown2005}. A non-deterministic Bell measurement
or modified version of the type-II fusion operation, $\rm B_{II}$,
identifies only two of the Bell states,
$\ket{H}\ket{V}\pm\ket{V}\ket{H}$, with success probability $1/2$.
It succeeds only when one detector from the upper two and another
from the lower two click at the same time.}\label{Btools}
\end{center}
\end{figure}

Four entangled coherent states, $\ket{\alpha}\ket{\alpha}\pm\ket{-\alpha}\ket{-\alpha}$ and
$\ket{\alpha}\ket{-\alpha}\pm\ket{-\alpha}\ket{\alpha}$, can be discriminated by coherent state Bell measurement, $\rm B_{\alpha}$, implemented by a 50:50 beam splitter and two photon number parity detectors (PNPD) as shown in Fig.~\ref{Btools} \cite{Jeong2001,Jeong2002QIC}. The four states after passing through the BS are
\begin{eqnarray}
\nonumber
\ket{\alpha}\ket{\alpha}+\ket{-\alpha}\ket{-\alpha}&\xrightarrow{\rm
BS}&\frac{1}{{\cal N}_e}\ket{\rm even}\ket{0},\\
\nonumber
\ket{\alpha}\ket{\alpha}-\ket{-\alpha}\ket{-\alpha}&\xrightarrow{\rm
BS}&\frac{1}{{\cal N}_o}\ket{\rm odd}\ket{0},\\
\nonumber
\ket{\alpha}\ket{-\alpha}+\ket{-\alpha}\ket{\alpha}&\xrightarrow{\rm
BS}&\frac{1}{{\cal N}_e}\ket{0}\ket{\rm even},\\
\nonumber
\ket{\alpha}\ket{-\alpha}-\ket{-\alpha}\ket{\alpha}&\xrightarrow{\rm
BS}&\frac{1}{{\cal N}_o}\ket{0}\ket{\rm odd},
\end{eqnarray}
where $\ket{\rm even}={\cal
N}_e\big(\ket{\sqrt{2}\alpha}+\ket{-\sqrt{2}\alpha}\big)$ and
$\ket{\rm odd}={\cal
N}_o\big(\ket{\sqrt{2}\alpha}-\ket{-\sqrt{2}\alpha}\big)$ (with
normalization factors ${\cal N}_e$ and ${\cal N}_o$) contain only even and
odd photon number states, respectively. Therefore, parity measurements on each
output mode enable one to discriminate between the four Bell states. A failure
occurs when no photon is detected in both the detectors due to the
nonzero overlap of $\bracket{0}{\pm\sqrt{2}\alpha}=e^{-\alpha^2}$.

When $\rm B_{\alpha}$ is performed on two hybrid qubits,
$\ket{\psi}=a\ket{+}\ket{\alpha}+b\ket{-}\ket{-\alpha}$ and
$\ket{\psi'}=a'\ket{+}\ket{\alpha}+b'\ket{-}\ket{-\alpha}$, coherent-state modes of two qubits are mixed by the 50:50 BS, such that the state evolves as
\begin{eqnarray}
\nonumber \ket{\psi}\ket{\psi'} &\xrightarrow{\rm
BS}& \frac{1}{2{\cal
N}_e}\big(aa'\ket{+}\ket{+}+bb'\ket{-}\ket{-}\big)
{\ket{\rm even}}\ket{0}\\
\nonumber &&+\frac{1}{2{\cal
N}_o}\big(aa'\ket{+}\ket{+}-bb'\ket{-}\ket{-}\big)\ket{\rm odd}\ket{0}\\
\nonumber &&+\frac{1}{2{\cal
N}_e} \big(ab'\ket{+}\ket{-}+ba'\ket{-}\ket{+}\big)\ket{0}{\ket{\rm
even}}\\
\nonumber &&-\frac{1}{2{\cal
N}_o}\big(ab'\ket{+}\ket{-}-ba'\ket{-}\ket{+}\big)\ket{0}\ket{\rm odd},
\end{eqnarray}
and the four possible states of remaining photon modes can be
discriminated from the results of the parity measurements on the output
modes of the BS. We note that its failure probability
\begin{eqnarray}
\nonumber
|\bra{0}\langle0\ket{\psi}\ket{\psi'}|^2 =
e^{-2\alpha^2}(|a|^2+|b|^2)(|a'|^2+|b'|^2) = e^{-2\alpha^2}
\end{eqnarray}
is lower than that obtained in CSQC
\cite{Jeong2001,Jeong2002QIC,Jeong2002,Ralph2003,Lund2008} due to
the orthonormality of hybrid qubit basis, {\it i.e.}, $|a|^2+|b|^2 = 1$.

\begin{figure*}
\includegraphics[width=0.85\linewidth]{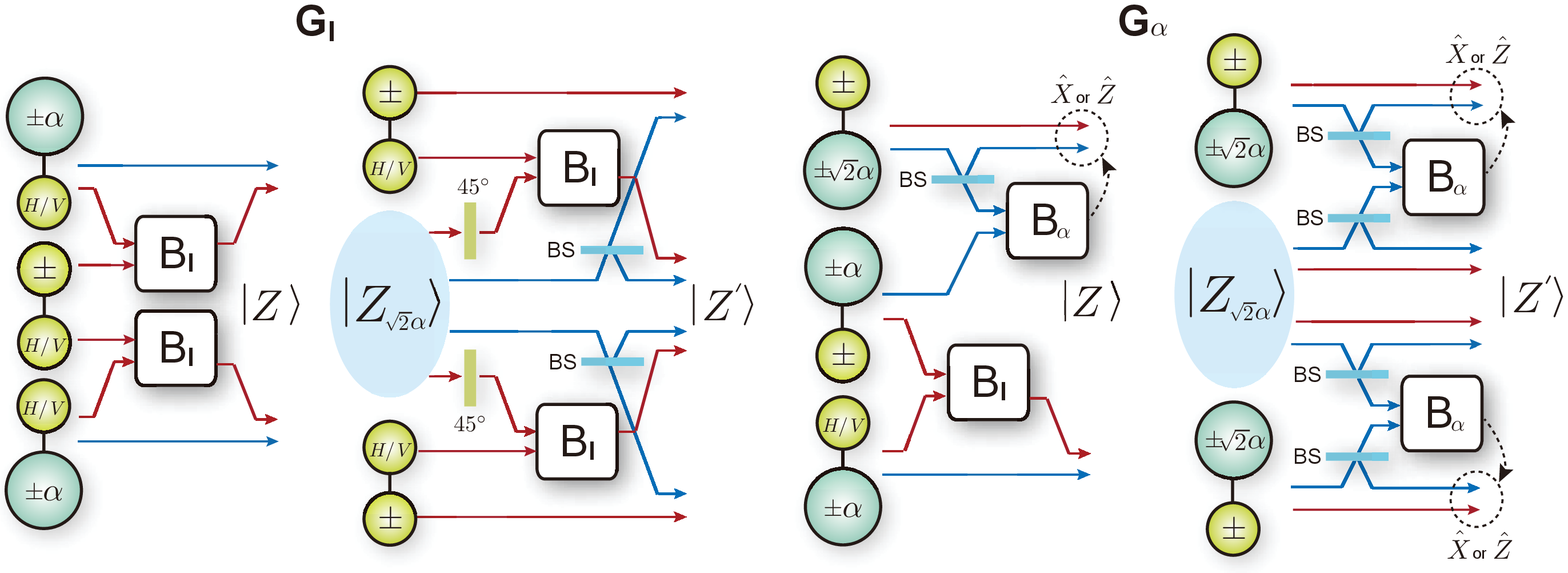}
\caption{Two schemes, $\rm G_I$ and $\rm G_\alpha$, for generating $\ket{Z}$ and $\ket{Z'}$
states. In a single-photon mode, $\pm$ and $H/V$ denote bases
$\{\ket{+},\ket{-}\}$ and $\{\ket{H}$,$\ket{V}\}$, respectively,
which can be modified by a polarization rotation. A Pauli-Z
operation (omitted in figure) is performed on the output qubit of
$\rm B_I$ when the measurement outcome of $\rm B_I$ is (H).
Likewise, for $\rm B_{\alpha}$ appropriate Pauli operations are
performed on the remaining qubit (denoted by dotted circle)
according to the measurement results as shown in
Table~\ref{tab:table1}.}\label{gene}
\end{figure*}


In the single-photon mode, two types of Bell measurements are used in Fig.~\ref{Btools}. A type-I fusion operation $\rm B_{I}$  \cite{DEBrown2005} performs a partial Bell measurement on the polarization states of photons. Its measurement outcome is either $H$ (when a photon is detected at the upper detector) or $V$ (when a photon is detected at the lower detector), which determines the operation actually carried out as: $\ket{+}\bra{H}\bra{H}-\ket{-}\bra{V}\bra{V}$ and $\ket{+}\bra{H}\bra{H}+\ket{-}\bra{V}\bra{V}$ for $H$ and $V$ clicks, respectively. It fails when two or no photon is detected at detectors, which
occurs with probability $1/2$.

A type-II fusion operation $\rm B_{II}$ \cite{DEBrown2005} performs
an incomplete Bell measurement with which only two out of the four
Bell states are distinguished \cite{Weinfurt1994,Lutkenhaus1999}. It
can be implemented by a polarizing beam splitter (PBS), wave plates
and photon detectors. It succeeds with probability $1/2$ when one
detector from the upper two and another from the lower two click at
the same time in Fig.~\ref{Btools} so that two Bell states can be
identified from the results: $\ket{H}\ket{V}-\ket{V}\ket{H}$ for the clicks $(H,H)$ or $(V,V)$, and $\ket{H}\ket{V}+\ket{V}\ket{H}$ for $(H,V)$ or $(V,H)$.

\renewcommand{\theequation}{B-\arabic{equation}}
\setcounter{equation}{0}
\setcounter{subsection}{0}
\section*{Appendix B: Generating entangled states}
\label{Appendix2}


A maximally entangled state of hybrid qubits $\ket{\Psi_C} \propto\ket{0_L}\ket{0_L}+\ket{1_L}\ket{1_L}$ can be generated by either of the two schemes, $\rm G_{I}$ or $\rm G_{\alpha}$, described in Fig.~\ref{fig3}. In $\rm G_{I}$, a hybrid pair with $\sqrt{2}$ times larger amplitude
$\ket{\psi_{\sqrt{2}\alpha}}=\ket{H}\ket{\sqrt{2}\alpha}+\ket{V}\ket{-\sqrt{2}\alpha}$ and a two-photon pair $\ket{H}\ket{+}+\ket{V}\ket{-}$ are merged by $\rm B_I$ operation:
\begin{eqnarray}
\nonumber
\Big(\ket{\alpha}\ket{\alpha}\ket{H}+\ket{-\alpha}\ket{-\alpha}\underbrace{\ket{V}\Big)
\Big(\ket{H}}_{\rm B_I}\ket{+}+\ket{V}\ket{-}\Big),
\end{eqnarray}
where the coherent-state mode was split into two modes by a 50:50 BS. Then, the
resulting state is given associated with the measurement outcome of $\rm B_I$ as
\begin{eqnarray}
\nonumber {\rm (H)~~Click~~}&:&~~
\ket{+}\ket{\alpha}\ket{+}\ket{\alpha}-\ket{-}\ket{-\alpha}\ket{-}\ket{-\alpha}\\
\nonumber {\rm (V)~~Click~~}&:&~~
\ket{+}\ket{\alpha}\ket{+}\ket{\alpha}+\ket{-}\ket{-\alpha}\ket{-}\ket{-\alpha},
\end{eqnarray}
and thus, by applying a Pauli-$Z$ operation on any qubit mode when the outcome is (H), we obtain $\ket{\Psi_C}$ whenever $\rm B_I$ operation succeeds with probability $1/2$.

In $\rm G_{\alpha}$, two $\ket{+}\ket{\alpha}\ket{\alpha}+\ket{-}\ket{-\alpha}\ket{-\alpha}$ states, obtained by applying a 50:50 BS to a hybrid pair with amplitude $\sqrt{2}\alpha$, are merged by $\rm B_{\alpha}$ operation. An appropriate Pauli operations dependent on the measurement outcome are applied on the remaining part so that the same state $\ket{\Psi_C}$ is produced whenever $\rm B_{\alpha}$ succeeds with probability $1-e^{-2\alpha^2}$ (see Table~\ref{tab:table1}). For example, when the measured outcome is (odd, 0), the resulting state is $\ket{+}\ket{\alpha}\ket{+}\ket{\alpha}-\ket{-}\ket{-\alpha}\ket{-}\ket{-\alpha}$
and a Pauli-Z operation on one qubit changes it to $\ket{\Psi_C}$.

\begin{table}[h]
\caption{\label{tab:table1}{Feed-forwards dependent on the results
of $\rm B_{\alpha}$}}
\begin{tabular}{|c|c|}
\hline
Measurement outcomes of $\rm B_{\alpha}$ & Pauli operations \\
\hline
(even, 0)& $\openone$ \\
\hline
(odd, 0) & $\hat{Z}$ \\
\hline
(0, even) & $\hat{X}$ \\
 \hline
(0, odd) & $\hat{Z}$ and $\hat{X}$ \\
 \hline
 (0, 0) & Failure \\
 \hline
\end{tabular}
\end{table}


A hybrid entangled state $\ket{Z}\propto\ket{0_L}\ket{0_L}+\ket{0_L}\ket{1_L}+\ket{1_L}\ket{0_L}-\ket{1_L}\ket{1_L}$
can be generated by either $\rm G_{I}$ or $\rm G_{\alpha}$ as shown in Fig.~\ref{gene}. In $\rm G_{I}$, one two-photon pair in $\ket{H}\ket{+}+\ket{V}\ket{-}$ and two hybrid pairs in $\ket{H}\ket{\alpha}+\ket{V}\ket{-\alpha}$ are merged by two $\rm B_I$ operations, and a Pauli-Z operation is applied on the outgoing mode of each $\rm B_I$ when the measurement outcome is (H). Then, $\ket{Z}$ is obtained when both $\rm B_I$ operations succeed with probability $1/4$. In $\rm G_{\alpha}$, three hybrid pairs are merged by $\rm B_I$ and $\rm B_{\alpha}$ as shown in Fig.~\ref{gene}. Here again a Pauli-Z operation is applied on the outgoing mode when the outcome of $\rm B_I$ is (H), and likewise appropriate Pauli operations (see Table~\ref{tab:table1}) are also applied after $\rm B_{\alpha}$ operation on the remaining qubit (denoted by dotted circle in Fig.~\ref{gene}). Thus the total success probability is given as $(1-e^{-2\alpha^2})/2$ since $\rm B_I$ and $\rm B_{\alpha}$ are used once each in the generation process.

A four-qubit entangled state
$\ket{Z'}\propto\ket{0_L}\ket{0_L}\ket{0_L}\ket{0_L}+\ket{0_L}\ket{0_L}\ket{1_L}\ket{1_L}
+\ket{1_L}\ket{1_L}\ket{0_L}\ket{0_L}-\ket{1_L}\ket{1_L}\ket{1_L}\ket{1_L}$
can also be generated by either $\rm G_{I}$ or $\rm G_{\alpha}$ as shown in Fig.~\ref{gene}. As shown in the Fig.~\ref{gene}, two photon-pairs and one $\ket{Z}$ state (with $\sqrt{2}$ times larger $\alpha$) are merged by two $\rm B_I$ operation in $\rm G_{I}$, while two hybrid-pairs and one $\ket{Z}$ state (all have $\sqrt{2}$ times larger $\alpha$) are merged by two $\rm B_{\alpha}$ operations in $\rm G_{\alpha}$. All $\rm B_I$ and $\rm B_{\alpha}$ are followed by appropriate Pauli operations dependent on their outcomes. The success probability of $\rm G_{I}$ is $(1/2)^4$ as it uses four $\rm B_I$ operations, while it is $(1-e^{-2\alpha^2})^3/2$ for $\rm G_{\alpha}$ that uses one $\rm
B_I$ and three $\rm B_{\alpha}$ operations.

\renewcommand{\theequation}{C-\arabic{equation}}
\setcounter{equation}{0}  
\section*{Appendix C: Error probabilities in lossy environment}
\label{NewAppendix}

In our numerical analysis, we consider errors caused by photon losses which is a major
obstacle to practical optical quantum computation.
The evolution of optical qubits in a lossy environment
can be described by solving a master equation $d\rho
/dt=\gamma(\hat{J}+\hat{L})\rho$ with
$\hat{J}\rho=\sum_i\hat{a}_i\rho\hat{a}_i^{\dag}$ and
$\hat{L}\rho=-\frac{1}{2}\sum_i(\hat{a}^{\dag}_i\hat{a}_i\rho+\rho\hat{a}^{\dag}_i\hat{a}_i)$
\cite{Phoenix1990}, where $\hat{a}_i(\hat{a}_i^{\dag})$ is the
annihilation (creation) operator for $i$-th mode. If the initial
state is a hybrid qubit
$\ket{\psi}=a\ket{+}\ket{\alpha}+b\ket{-}\ket{-\alpha}$, it evolves
into a mixed state,
\begin{widetext}
\begin{equation}
\label{eq:master} 
\ket{\psi}\xrightarrow{\eta}(1-\eta)\bigg(\frac{1+e^{-2\eta\alpha^2}}{2}\ket{\psi'^{+}}\bra{\psi'^{+}}
+\frac{1-e^{-2\eta\alpha^2}}{2}\ket{\psi'^{-}}\bra{\psi'^{-}}\bigg)
+\eta\bigg(\frac{1}{2{\cal
N}^{+2}_{a,b}}\ket{\phi^{+}}\bra{\phi^{+}}+\frac{1}{2{\cal
N}^{-2}_{a,b}}\ket{\phi^{-}}\bra{\phi^{-}}\bigg),
\end{equation}
\end{widetext}
where the loss rate is defined as $\eta= 1-e^{-\gamma t}$. Here
$\ket{\psi'^{\pm}}=a\ket{+}\ket{\alpha'}\pm b\ket{-}\ket{-\alpha'}$
with $\alpha' = \sqrt{1-\eta}\alpha$ are possible resulting states
of a hybrid qubit when only the amplitude of the coherent state is reduced,
while $\ket{\phi^{\pm}}={\cal
N}^{\pm}_{a,b}|0\rangle(a\ket{\alpha'}+b\ket{-\alpha'})$ with normalization
factors ${\cal N}^{\pm}_{a,b}$ are possible remaining coherent
states when loss occurs in the single-photon mode. The loss rate $\eta$ is considered
a known value and the states
$\ket{\psi'^{+}}$ and $\ket{\phi^{+}}$ do not contain any logical
errors. On the other hand, $\ket{\psi'^{-}}$ and $\ket{\phi^{-}}$ contain Pauli-$Z$
errors. Note that the probabilities of the states $\ket{\phi^{\pm}}$
depend on $a$ and $b$ since coherent state qubits are carrying
information in a nonorthogonal basis. We here choose the worst case of the
error rate, which is obtained from the condition of the minimum value of
${\cal N}^{-}_{a,b}$. This results in ${\cal N}^{\pm}_{a,b}=1$ and
the state~(\ref{eq:master})  becomes
\begin{equation}
\label{eq:master2}
\begin{aligned}
&\bigg((1-\eta)\frac{1+e^{-2\eta\alpha^2}}{2}\ket{\psi'^{+}}\bra{\psi'^{+}}+\frac{\eta}{2}\ket{\phi^{+}}\bra{\phi^{+}}\bigg)\\
&~~~~~~~+\bigg((1-\eta)\frac{1-e^{-2\eta\alpha^2}}{2}\ket{\psi'^{-}}\bra{\psi'^{-}}+\frac{\eta}{2}\ket{\phi^{-}}\bra{\phi^{-}}\bigg)_{\rm
Z},
\end{aligned}
\end{equation}
where the last two terms represent the states conveying Pauli-Z
errors.

Based on the  above result, we can investigate all possible errors
caused by losses in our scheme. The failure probability of
teleportation,
$P_f$ in Eq.~(1), is
changed  to $P_f'$ in Eq.~(2) by losses.
The modified error probability $P_f'$ is
a weighted sum of the error probabilities
obtained
using the loss rate $\eta$ for the photon in the single-photon mode and the reduced
amplitude of the coherent state $\alpha'$.
The component in the second term, $2/(1+e^{2\alpha'^2})$,
corresponds to the failure probability of the coherent-state teleportation in the presence
of loss obtained in Ref.~\cite{Lund2008}.

If a Hadamard or CZ gate fails, this means that a hybrid-Bell
measurement failed (or both the hybrid-Bell measurements failed in
the case of a CZ gate).
In this case, it can be shown that the output qubit(s) experiences depolarization
and become fully mixed. This can be modeled in our scheme by
applying a random Pauli operation to the qubit, {\em i.e.} $Z$ and $X$
Pauli errors occur independently with equal probabilities.

The loss in a photon part can be detected whenever performing a $\rm B_{II}$
operation in teleportation and the loss is compensated once the
teleportation succeeds ({\it i.e.}, if either $\rm B_{II}$ or $\rm B_\alpha$
is successful done in the hybrid-Bell measurement).
If photon loss at a photon part (which occurs with probability $\eta$)
is noticed, it means that a Pauli $Z$ error might have occurred with probability 1/2 as implied in Eq.~(2).

We also consider errors that may occur in quantum memory that used to
store qubits that are not undergoing gate operations.
In quantum memory, losses in either photon or coherent-state mode
induce Pauli-$Z$ errors with the rate
\begin{eqnarray}
\nonumber p &=&
(1-\eta)\frac{1-e^{-2\eta\alpha^2}}{2}+\frac{\eta}{2}=
\frac{1}{2}\big\{1-(1-\eta)e^{-2\eta\alpha^2}\big\},
\end{eqnarray}
which is obtained by summing the probabilities of the last two terms in
Eq.~(\ref{eq:master2}).

The entangled states $|Z\rangle$ and $|Z'\rangle $ are
necessary resources for Hadamard and CZ gates in our scheme.
Losses in the generation processes of $\ket{Z}$ and $\ket{Z'}$ may
cause errors in output qubits of Hadamard and CZ gates. We consider
these errors by assuming that losses occur immediately after the
resource states are produced \cite{Dawson2006,Lund2008}. In this
model, photon loss at one qubit in $\ket{Z}$ induces a Pauli-$Z$
error, while loss at the other qubit induces a Pauli-$X$ error in a teleportation process
for a Hadamard gate.
The error probability for both Pauli-$Z$ and $X$ errors is $p$, which obtained
exactly in the same way as the one obtained in quantum memory  from Eq.~(2).
Losses in any
qubit of $\ket{Z'}$ induces a Pauli-$Z$ error in a CZ gate with
probability $p$ per qubit.

The preparation of the diagonal qubits ({\em i.e.} hybrid pairs) are required
for the telecorrection protocol.
We also consider possible errors in the
preparation of a diagonal qubit by assuming
that loss occurs immediately after its generation so that it may
convey a Pauli-$Z$ error with probability $p$. It is also assumed that
there is no additional error caused by $X$-basis measurement used in
the telecorrection protocol.

\renewcommand{\theequation}{D-\arabic{equation}}
\setcounter{equation}{0}  
\section*{Appendix D: Effects of multiphoton contributions}
\label{appendix:multi}

Due to experimental imperfections, there could be more than one photon in one part of a two-photon pair (or in the single-photon part of a hybrid pair). First, in the off-line preparation process using $\rm B_I$, multiphotons in the single photon part can be partially detected (for example, by the case when detectors simultaneously click in a $\rm B_I$ operation) and such cases can simply be discarded. Such errors only result in a slight increase of the resource requirement given that the multi-portion contribution is slight compared to the single-photon contribution.

First, we point out that roughly half of the multiphoton contributions will be discarded by $\rm B_I$ during the generation process. To explain this, we use an approximate approach: let us assume that the generated state (for the case of a two-photon pair but the same analysis apply a hybrid pair) is
$$
\propto|H\rangle|H\rangle+|V\rangle|V\rangle+\lambda(|2H\rangle|2H\rangle+|2V\rangle|2V\rangle)
$$
where $|2H\rangle$ ($|2V\rangle$) is a two-photon state with the horizontal (vertical) polarization and  $\lambda$ is assumed to be a small value.
Considering a typical down-conversion process, we can assume that $\lambda$ is very small and the probability of having three photons (or more) in one mode is negligible \cite{Kwiat1995,our-two}. We shall also ignore $\lambda^2$ factors in the following calculations because $\lambda$ is already very small. Due to the symmetry, it is sufficient to consider the following four possible possibilities (out of total eight), $|2H\rangle|V\rangle$ or $|2V\rangle|V\rangle$ or
$|2H\rangle|H\rangle$ or $|2V\rangle|V\rangle$ (upper and lower modes in order), as the input to the PBS of the $\rm B_I$ process shown in Fig.~\ref{Btools}.

The two PBSs used for a $\rm B_I$ process are assumed to pass ``H'' polarizations and to reflect ``V''. Note also that detectors used for a standard $\rm B_I$ process requires single-photon detectors that discriminate between 0, 1 and more than 1 photons \cite{DEBrown2005}. For the first two cases, $|2H\rangle|V\rangle$ or $|2V\rangle|V\rangle$, all the photons go to either the upper direction together or the lower by the first PBS. These cases are all failures and do not make any difference from those without multiphoton contirubutions. For the third case, $|2H\rangle|H\rangle$, one photon goes to the upper part and the two photons go to the lower part. This result is considered a ``success'' and the two-photon contribution goes into the inline process. For the fourth case, $|2V\rangle|V\rangle$, the two photons go to the upper part and one photon to the lower part. The two detectors then recognize that there exist more than one photon. (Note that even though the detectors used for the ${\rm B_I}$ operation discriminates between a single photon and “two or more photons”, it cannot resolve more than two photons.) This case is a failure and simply discarded. Therefore, we can conclude that the multiphoton contributions are reduced to about half by the ${\rm B_I}$ operations during the resource-generation stage (with a slight increase of the resource cost).

Next, for the remaining multiphoton contributions in the in-line process, we can make the same analysis as above with the four cases for the two input modes. It should be noted that a standard $\rm B_{II}$ operation use four {\it on/off} detectors, and two of those detectors click when the result is a success. If the ``surplus'' photon goes to one of those two detectors, the result is considered as a success, but success events possibly convey unlocatable Pauli-Z error with a probability $1/2$. 
On the other hand, if the ``surplus'' photon goes to one of the other two detectors, the result is a failure {\em i.e.} a locatable error. For example, if the input state is $|2H\rangle|V\rangle+|2V\rangle|H\rangle$, all possible cases for the success events in front of the second PBSs are $\ket{H}\ket{2H}$,  $\ket{2H}\ket{H}$, $\ket{V}\ket{2V}$, $\ket{2V}\ket{V}$, $\ket{H}\ket{2V}$, $\ket{2H}\ket{V}$, $\ket{V}\ket{2H}$ and $\ket{2V}\ket{H}$. If $(H,V)$ or $(V,H)$ clicks occur  at the detectors, with probability $1/2$, the result is a 
``correct'' success
and and the multiphoton contribution will simply disappear at the detectors.
However,  if $(H,H)$ or $(V,V)$ events occurs, the result is an ``incorrect'' success and it will deliver an unnoticed Pauli-$Z$ error to the teleported qubit. All other cases in front of the second PBSs such as $\ket{H}\ket{HV}$,  $\ket{HV}\ket{H}$, $\ket{V}\ket{HV}$, and $\ket{HV}\ket{V}$ are detected as failures.

Finally, when there is photon loss, there is some additional possibility of errors caused by the multi-photon contributions. For example, the multiphoton contribution may cause a ``click at a wrong detector'' and a  ``photon missing at a correct detector.'' This type of error cannot be noticed and thus an unlocatable error.

In summary, the multiphoton contributions will play a role to increase the resource cost and decrease the noise threshold. However, based on the discussions above, we can expect that the effects of multiphoton contributions are relatively very small because most of them would have been discarded by $\rm B_I$ or detected by $\rm B_{II}$
during the generation and inline processes.
We also point out that, of course, such multiphoton effects, which are not considered in most of the references, are present not only in our scheme but also in any LOQC-type approaches (including pLOQC) where two-photon pairs, generated by the parametric down conversion, are used as resources \cite{boook}.

\end{document}